\documentclass{iau}
\usepackage{graphicx}

\newcommand{\apj}{\textit{ApJ}}				
\newcommand{\aap}{\textit{A\&A}}				
\newcommand{\mnras}{\textit{MNRAS}}			
\newcommand{\sci}{\textit{Science}}				
\newcommand{\sphy}{\textit{Solar Phys.}}			

\title[Starspot modeling] 
{Stellar Magnetism and starspots: \\ the implications for exoplanets}

\author[Conrad Vilela, John Southworth \& Carlos del Burgo]   
{Conrad Vilela$^{1,\star}$, John Southworth$^1$ \and Carlos del Burgo$^2$}

\affiliation{$^1$Astrophysics Group, Keele University, \\ 
Keele, Staffordshire, ST5 5BG \\
$^2$Instituto Nacional de Astrof\'\i sica, \'Optica y Electr\'onica, \\
Luis Enrique Erro 1, Sta. Ma. Tonantzintla, Puebla, Mexico \\
$\star$email: {\tt c.vilela@keele.ac.uk}}

\pubyear{2013}
\volume{302}  
\pagerange{119--123}
\jname{Magnetic Fields Throughout Stellar Evolution}
\editors{M. Jardine, P. Petit \& H. Spruit.}
\begin{document}

	\maketitle

	\begin{abstract}
		Stellar variability induced by starspots can hamper the detection of exoplanets and bias planet property estimations. These features can also be used to 
		study star-planet interactions as well as inferring properties from the underlying stellar dynamo. However, typical techniques, such as ZDI, are not possible for most
		host-stars. We present a robust method based on spot modelling to map the surface of active star allowing us to statistically study the effects and interactions
		of stellar magnetism with transiting exoplanets. The method is applied to the active Kepler-9 star where we find small evidence for a possible interaction between 
		planet and stellar magnetosphere which leads to a 2:1 resonance between spot rotation and orbital period.
	\end{abstract}

	\firstsection 
	\section{Introduction}
		Cool stars (F-M dwarfs), mainly solar-like stars, have partially convective interiors -- a radiative core and a convective envelope. The turbulent cyclonic motion
		of the plasma in the convective layer generates an underlying magnetic dynamo which drives a variety of stellar features. The magnetic fields 
		produced penetrate the stellar atmosphere from photosphere to corona forming starspots, plages, loops, flares, etc 
		(\cite[Berdyugina 2005]{Berdyugina05}). These features can be observed with a variety of techniques. The more popular are 
		Zeeman-Doppler imaging (ZDI) and spectropolarimetry which provide information on the topology of the magnetic field 
		(\cite[Donati \& Semel 1990]{DS90}; \cite[Phan-Bao \etal\ 2009]{PB09}). However, these techniques 
		become ill-suited for faint stars and slow rotators. With the increasing number of solar-like host-stars discovered by the {\emph Kepler} satellite we need to 
		look at different ways of studying the stellar magnetic topology and its effects on planetary companions. 
		
		One can use activity proxies, such as the Ca II H\&K line inversion ($S_{HK}$ or $\log{R'_{HK}}$) which trace chromospheric activity
		(\cite[Wilson 1978]{Wilson78}, \cite[Noyes \etal\ 1984]{Noyes84}). Using this activity proxy \cite[Knutson \etal\ (2010)]{Knutson10} found that exoplanets with no 
		temperature inversion are more likely to orbit active stars as these produce higher  UV-fluxes which photo dissociate molecular absorbers (e.g., TiO and VO). 
		A more common and accessible technique, based on photometry, is the use of rotational modulated light curves caused by the presence of starspots. These surface 
		inhomogeneities are formed by local magnetic fields suppressing the vertical heat transport in the convective layer. They are cooler (1000-3000 K) than their
		surrounding photosphere and are carried through the stellar surface by stellar rotation.
				
		In the exoplanet community starspots are considered a source of noise as they can hamper the detection and characterisation of the 
		companion planet. If present during transit they can bias the determination of the planet radius and density (\cite[Czesla \etal\ 2009]{Czesla09}). The transit 
		shape is also affected, flattening the trough and widening the transit, blurring the transit time and duration (\cite[Oshagh \etal\ 2013]{Oshagh13}). On the other 
		hand starspots can also be used as a characterisation tool revealing properties of the system, such as differential rotation rates, rotation-age relations or 
		star-planet interactions. \cite[Sanchis-Ojeda \& Winn (2011)]{SO11)} used occulted spots to provide the obliquity and misalignment of the WASP-4 system. 
		Close-in planets can become circularised where orbital periods are synchronised with stellar rotation due to the interaction with the magnetosphere 
		(\cite[Laine \etal\ 2008]{Laine08}). Furthermore, this can induce changes in the magnetic topology and tidal interactions which can produce a spin-up of the 
		stellar rotation (\cite[Brown \etal\ 2011]{Brown11}).
		
		The aim is to produce a robust methodology to model rotational modulated light curves and statistically analyse the effects and interactions of stellar magnetism 
		and exoplanet companions. The following sections describe the method used to map the surface of active stars and the preliminary results when applied to a 
		discovered \emph{Kepler} system -- Kepler-9.

		\begin{table}[h!]
			\begin{center}
				\caption{Definition of spot parameters, description and fitting constraints}
				\label{table1}
				{\scriptsize
					\begin{tabular}{lcccc}
						\hline
						Parameters & & Description & & Constraints \\
						\hline
						$U$ & & stellar unspotted flux & & min -- max amplitude\\
						$i_{axis}$ & & stellar rotational axis inclination & & $0^{\circ}$ -- $90^{\circ}$ \\
						$\mu$ & & linear limb darkening coefficient & & fixed\\
						$\kappa$ & & spot contrast & & 0 -- 1 \\
						$\theta$ & & latitude of spot centre & & $-90^{\circ}$ -- $90^{\circ}$ \\
						$\psi^{a}$ & & longitude of spot centre & & fixed \\
						$\gamma$ & & spot angular radius & & $0^{\circ}$ -- $90^{\circ}$ \\
						$E$ & & spot epoch & & min -- max timestamp \\
						$P_{rot}^{b}$ & & spot rotational period & & $P_{rot}/2$ -- $2P_{rot}$ \\
						\hline
					\end{tabular}
				}
			\end{center}
			
			\scriptsize{
				\hspace*{20mm}
				{\it Notes:} \\ 
				\hspace*{22mm} 
				$^a$ Spot position is better determined by $E$ (See \cite[Croll \etal\ (2006)]{Croll06}). \\ 
				\hspace*{22mm}
				$^b$ The initial values are obtained from {\sc SigSpec} and left to vary.
			}
		\end{table}

	\section{Methodology}
		Spot modelling techniques are based on reproducing the rotational modulated light curve into a physical map of the starspot coverage. The main 
		disadvantage of these techniques is the high degeneracy and non-uniqueness between the model parameters.  Various approaches have been used to solve 
		this, such as a continuos spot distribution or a fixed number of spots (\cite[Lanza \etal\ 2007]{Lanza07}). However, these approaches either limit the level of activity 
		observed or cannot precisely extract spot properties. In practice, one would like to find a unique un-biased solution to the problem without any loss of information. 
		To achieve this the method presented is divided into two steps, a frequency analysis and spot modelling.
			
		{\underline{\it Frequency analysis}}.  Spot $P_{rot}$ can be easily obtained from the stars light curve by analysing the signal frequencies from the rotational
		modulations. To achieve this we implement a statistical technique based on the Lomb-Scargle periodogram, Significance Spectrum ({\sc SigSpec}) 
		(\cite[Reegen 2007]{Reegen07}). The frequencies computed by {\sc SigSpec} are constrained to the range $0.6-0.016~cycles/day$ 
		($P_{rot} \approx 1.6 - 60~days$). The upper limit ensures that {\sc SigSpec} does not detect frequencies which might correspond to background noise, 
		as Solar-like stars tend to be slow rotators. The lower limit ensure that the detected frequency is representative of a full observed cycle, i.e., the same spot has 
		to be in view at least twice within the light curve. This avoid the detection of frequencies belonging to spot or magnetic cycles and long trends. As an additional
		functionality {\sc SigSpec} can also determine the corresponding phase of the frequency, constraining the spots $E$. 
		   
		{\underline{\it Spot modelling}}. To model the rotational modulation we use the 4-term limb darkening spot model from \cite[Kipping (2012)]{Kipping12}. The model
		has 7 parameters described in table\,\ref{table1}, of which $\psi$ is calculated using the spots' $E$. To reduce computational time and avoid introducing
		degeneracies the models are computed without spot evolution, migration or differential rotation. However, in general these properties can be estimated from 
		the spots themselves. Although $P_{rot}$, $E$ and to a certain degree $\theta$ are constrained, $\kappa$ and $\gamma$ are highly correlated and typical 
		optimisation techniques will not provide a global unique solution. To overcome this degeneracy the light curve is fitted using the {\sc MultiNest} code based on 
		Bayesian statistics and elliptical decomposition (\cite[Feroz \etal\ 2013]{Feroz13}). Given the Bayesian nature of {\sc MultiNest}  we can also obtain the marginalising
		integral or model evidence used for model selection through Bayes factor ($B$). As such we iterate through all frequencies obtained from {\sc SigSpec}. At each 
		iteration a spot is added to the model until the log difference in $B$ ($K$) between consecutive iterations is less than 2, showing no strong evidence for the current
		model (\cite[Jeffreys 1961]{Jeffreys61}).  

	\section{Results from Kepler-9}
		Kepler-9 is a G type stars with 3 transiting planets observed by \emph{Kepler} for $\sim3$ years. \cite[Holman \etal\ (2010)]{Holman10} reported this star to have
		photometric variations slightly larger than the Sun and an inversion of the Ca II H\&K line core indicative of moderate activity. The similarities between Kepler-9 
		and the Sun make this a perfect target for the analysis. We only use the first 7 quarters (Q1-Q7) of the \emph{Kepler} data fitting each individual quarter.
		
\begin{figure}[h!]
			\begin{center}
				\includegraphics[width=\textwidth]{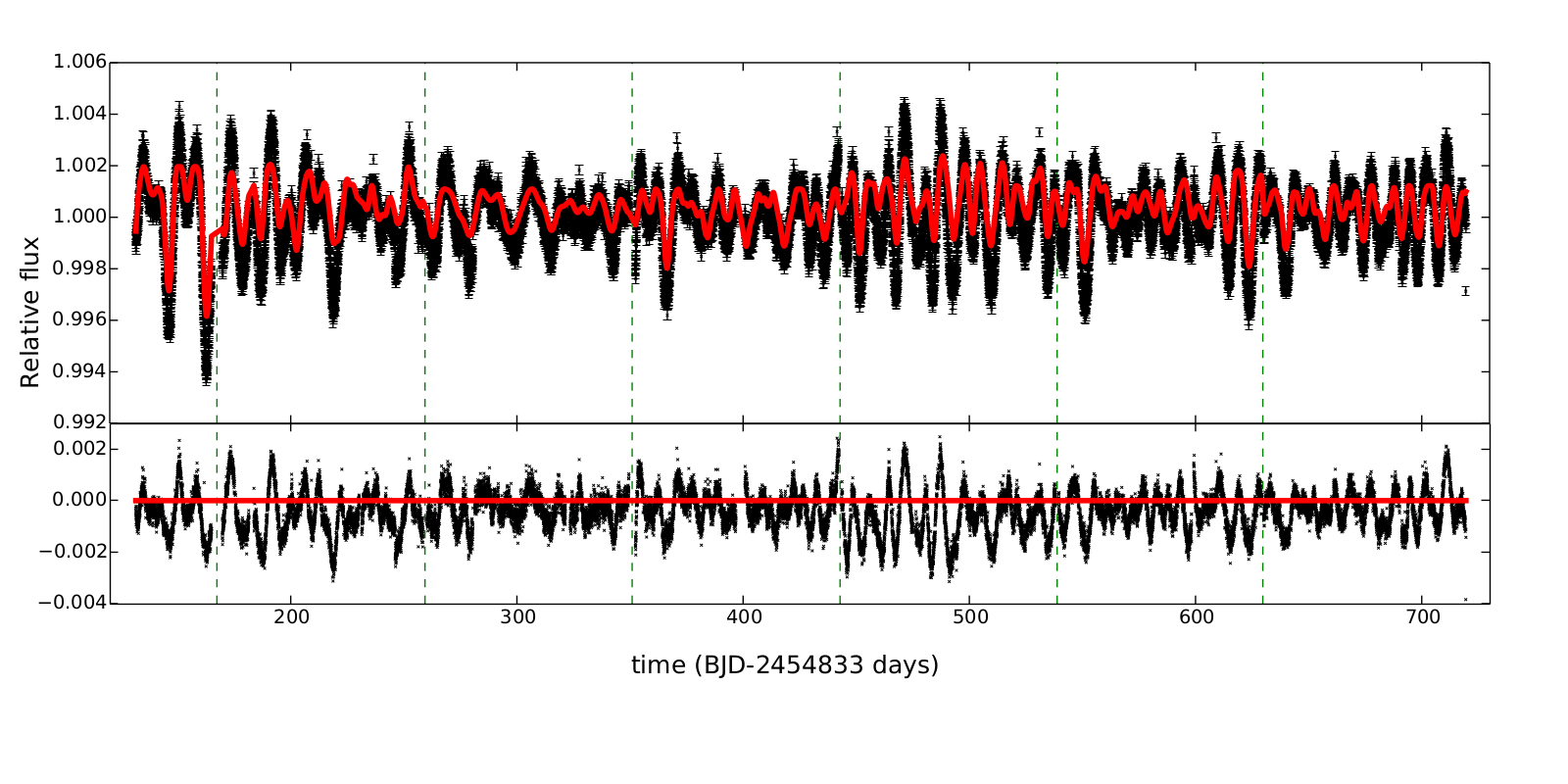}
				\vspace*{-1.5cm}
				\caption{{\it Top}: Kepler-9 combined light curve (Q1-Q7) with the transits from both planets removed. Black points represent the PDC-MAP data
						from \emph{Kepler} and the solid line is the best spot model fit for each of the quarters. The vertical dashed lines separate each quarter.
						{\it Bottom}: Residual from each quarter from the best spot model fit}
				\label{fig1}
			\end{center}
		\end{figure}

		We are able to reproduce the Kepler-9 light curve seen in Fig.\,\ref{fig1}. The residuals show a periodic signal which could be signs of more spots present than 
		those fitted by the models. However, the small value of $K$ ($< 0 $) indicates that fitting more spots, will neither provide a better fit or more information. Hence, 
		this periodicity is interpreted as a combination of spot evolution and migration. Fig.\,\ref{fig1} also shows a change in both amplitude and shape of the modulation 
		with time, more precisely between quarters, reproduced by the model (solid line in Fig.\,\ref{fig1}). This is interpreted as a change in spot number from quarter to
		quarter indicative of a possible spot cycle ($> 90$ days) similar to that seen in the Sun. As the models do not include spot evolution, migration or differential
		rotation, the apparent spot cycle further supports the idea that the periodicity in the residuals is due to spot evolution and migration. 
		
		The resulting best fit spot model shows that the vast majority of the rotational modulation is produced by large near-polar spots ($\theta \approx -85^{\circ} to
		-75^{\circ}$, $\gamma \approx 50^{\circ} - 90^{\circ}$) which could correlate with open magnetic field lines. We infer from this that the large near-polar spots are the
		projection of coronal holes on the photosphere. However, we cannot resolve the stellar surface and, thus, unable to distinguish between a large spot and a group 
		of smaller spots. The alternative explanation is that the large spots ($\gamma > 50^{\circ}$) correspond to spot groups. If these spots are evolving it is very likely 
		due to smaller spots disappearing within the spot group. This would mean that the large spot would be tracing active regions with complex coronal structures.

		Lastly the best fit spot model shows spot $P_{rot}$ in a possible near 2:1 resonance with the planet's orbital period. We find that spots close to the pole have 
		$P_{rot}$ (8 - 12 days) near half the orbital period of planet b ($\sim$9.6 days). Spots near the equator show $P_{rot}$ (15 - 22 days) that are closer to half the
		orbital period of planet c ($\sim$19.4 days). This could infer a possible star-planet interaction which can be interpreted as different evolutionary paths for both
		planets. However, we are unable to identify if the interaction is magnetic or tidal in origin. Furthermore, both planets are close enough to the host-star 
		($a \approx 0.1-0.2 au$) to have an effect on the magnetic topology of the star.
		
\end{document}